\documentclass[prd,twocolumn,showpacs,showkeys,%
  preprintnumbers,amsmath,amssymb,floatfix]{revtex4}

\usepackage{graphicx,color}
\usepackage{dcolumn}
\usepackage{bm,slashed}

\def\beq{\begin{eqnarray}}
\def\eeq{\end{eqnarray}}

\begin{document}

\preprint{SI-HEP-2013-18, QFET-2013-14, TTK-14-05, SFB/CPP-14-07, TUM-HEP-928/14}

\title{$B$-Meson Light-Cone Distribution Amplitude:
\\ Perturbative Constraints and Asymptotic Behaviour in Dual Space}

\author{Thorsten Feldmann}
\email{thorsten.feldmann@uni-siegen.de}
\affiliation{%
Theoretische Elementarteilchenphysik,
Universit\"at Siegen, 
57068 Siegen, Germany}

\author{Bj\"orn O.~Lange}
\email{lange@tp1.physik.uni-siegen.de}
\affiliation{%
Theoretische Elementarteilchenphysik,
Universit\"at Siegen, 
57068 Siegen, Germany}

\author{Yu-Ming Wang} \email{yuming.wang@tum.de}
\affiliation{%
Institut f\"ur Theoretische Physik~E, 
RWTH Aachen, 
52056 Aachen, Germany; 
\\
Technische Universit\"at M\"unchen, 
Physik Department, 
85747 Garching, Germany}

\date{\bf 4.~April 2014}%

\begin{abstract}
Based on the  dual
representation in terms of the recently established 
eigenfunctions  of the evolution kernel
in heavy-quark effective theory,
we investigate the description of the 
$B$-meson light-cone distribution amplitude (LCDA)
beyond tree-level. In particular, in 
dual space, small and large momenta do not mix
under renormalization, and therefore perturbative
constraints from a short-distance expansion in
the parton picture can be implemented independently
from non-perturbative modelling of long-distance effects.
It also allows to (locally) resum perturbative
logarithms from large dual momenta at fixed values of the
renormalization scale.
We construct a generic procedure to combine
perturbative and non-perturbative information on the $B$-meson
LCDA and compare different model functions and the resulting
logarithmic moments which are the relevant hadronic parameters
in QCD factorization theorems for exclusive $B$-meson decays.
\end{abstract}



\maketitle


\section{Introduction}

In a recent paper \cite{Bell:2013tfa}, it has been shown that
the evolution kernel \cite{Lange:2003ff}, which determines
the 1-loop renormalization-group (RG) evolution of the $B$-meson
light-cone distribution amplitude (LCDA) in heavy-quark effective theory
(HQET), can be diagonalized 
by an appropriate integral transform. As the so-defined new function
(dubbed ``spectral'' or ``dual'' in \cite{Bell:2013tfa}) renormalizes
\emph{locally} with respect to its argument (denoted as $\omega'$ in
the following), the properties at large and small values of $\omega'$
are clearly separated. In particular, we expect that for values of
$\omega'$ much larger than the typical hadronic scale the dual
function can be constrained by perturbative physics related to
the operator product expansion (OPE) in the heavy-quark limit
\cite{Lee:2005gza}. On the other hand, the behaviour at 
small values of $\omega'$ could be adjusted to results from non-perturbative 
QCD methods.
Finally, the experimental results for exclusive $B$-decay observables
constrain the logarithmic moments of the dual LCDA in QCD factorization
theorems (see \cite{Beneke:1999br} and related work).
In this way, one can construct parametrizations for
the $B$-meson LCDA which include constraints from short- and long-distance
theoretical predictions and experimental information simultaneously 
(for similar ideas in a
different context see also  \cite{Ligeti:2008ac} or \cite{Ball:2005ei}).

Our paper is organized as follows. In Sec.~\ref{sec:2} we review the
diagonalization of the renormalization kernel for the LCDA, which
allows to describe the perturbative and non-perturbative domains of
the dual function separately. In Sec.~\ref{sec:3} we discuss the
perturbative information available from regularized moments of the LCDA and
translate them to the dual function. The main topic of this paper,
to wit the construction of the dual LCDA from a given model ansatz
while respecting all known perturbative constraints, is addressed in
Sec.~\ref{sec:4}, and particularly in
Eq.~(\ref{ansatz}) below. In Sec.~\ref{sec:4a} 
we discuss the logarithmic moments of the dual function, with particular
focus on the contributions from small and large values of $\omega'$.
Illustrative examples and
their logarithmic moments are discussed in Sec.~\ref{sec:5}, followed
by a summary and an appendix with mathematical details.


\section{Diagonalization of the Kernel \label{sec:2}}

The leading LCDA of the $B$-meson in HQET, 
which is denoted as $\phi_B^+(\omega)$ in this work, is defined 
from the matrix element of a 2-particle light-cone operator \cite{Grozin:1996pq},
\begin{eqnarray}
&& \tilde f_B m_B \, \phi_B^+(\omega) = \int \frac{d\tau}{2\pi} \, e^{i\omega\tau} 
 \cr 
&& {} \times    \langle 0 | \bar q(\tau n) \, [\tau n,0]\, \slashed n \gamma_5 \, h_v(0)| \bar B(m_B v)\rangle
\label{Bdef}
\end{eqnarray}
where $n^\mu$ is a light-like vector, $[\tau n,0]$ is a gauge-link represented
by a Wilson line in the $n^\mu$-direction, 
and $\tilde f_B$ is the $B$-meson decay constant in HQET. The variable $\omega$ 
represents the $n$-projection of the light quark's momentum.

The renormalization of the non-local light-cone operator in the presence of a
static heavy quark in HQET induces a particular
renormalization-scale dependence. 
This gives rise to the Lange-Neubert (LN) kernel entering 
the renormalization-group equation (RGE) \cite{Lange:2003ff}
\begin{eqnarray}
 \frac{d}{d\ln\mu} \, \phi_B^+(\omega) &=&- \left[\Gamma_{\rm cusp} \, \ln \frac{\mu}{\omega} 
  + \gamma_+ \right] \phi_B^+(\omega) 
  \cr && {} - \omega \, \int\limits_0^\infty d\eta \, \Gamma(\omega,\eta) \, \phi_B^+(\eta) \,.
  \label{eq:rge1}
\end{eqnarray}
The leading terms in the various contributions to the anomalous dimension 
in units of $\frac{\alpha_s}{4\pi}$ are
\begin{eqnarray}
&& \Gamma_{\rm cusp}^{(0)} \equiv \Gamma_0 =4 \, C_F \,, \qquad 
\gamma_+^{(0)} \equiv \gamma_0 = -2 \, C_F \,, 
\cr 
&& 
\Gamma^{(0)}(\omega,\eta) = - \Gamma_0 \left[ \frac{\theta(\eta-\omega)}{\eta(\eta-\omega)}
+ \frac{\theta(\omega-\eta)}{\omega(\omega-\eta)} \right]_+ 
 \,.
\end{eqnarray}
As shown in \cite{Lee:2005gza}, the explicit solution for $\phi_B^+(\omega,\mu)$ 
can be written in closed form as a convolution integral 
involving hypergeometric functions.
(If not otherwise stated, in the following the
renormalization-scale dependence is implicitly understood, i.e.\ 
$\phi_B^+(\omega) \to \phi_B^+(\omega,\mu)$ etc. Similarly, the
anomalous dimensions have a perturbative expansion in the strong coupling,
$\Gamma_{\rm cusp}=\Gamma_{\rm cusp}(\alpha_s(\mu))$ etc.) 

In a recent article \cite{Bell:2013tfa} some of us have shown that 
the solution of the RG equation simplifies when the LCDA is 
represented in a  ``dual'' momentum space, defined 
via
\begin{eqnarray}
\phi_B^+(\omega) &=& 
\int\limits_0^\infty \frac{d\omega'}{\omega'} \, 
\sqrt{\frac{\omega}{\omega'}} \, J_1\left( 2 \, \sqrt{\frac{\omega}{\omega'}}\right)  \rho_B^+(\omega') \,,
\label{trans}
\end{eqnarray}
where $\rho_B^+$ defines a spectral function in the dual variable $\omega'$,
and $J_1(z)$ is a Bessel function. (A similar relation holds for the other 2-particle LCDA $\phi_B^-(\omega)$,
see \cite{Bell:2013tfa}, 
which reproduces the corresponding RGE in the Wandzura-Wilczek approximation 
\cite{Bell:2008er,Knodlseder:2011gc,DescotesGenon:2009hk}.)
The inverse transformation analogously reads
\begin{eqnarray}
\rho_B^+(\omega') &=&
 \int\limits_0^\infty \frac{d\omega}{\omega} \, \sqrt{\frac{\omega}{\omega'}} \,
 J_1\left( 2 \, \sqrt{\frac{\omega}{\omega'} }\right) \phi_B^+(\omega) \,.
\end{eqnarray}
The dual function $\rho_B^+(\omega')$ now obeys the simple RGE
\begin{eqnarray}
\frac{d\rho_B^+(\omega')}{d\ln\mu}
 &=& - \left[\Gamma_{\rm cusp} \, \ln \frac{\mu}{\hat\omega'} + \gamma_+ \right]
 \rho_B^+(\omega') \,,
 \label{eq:rge2}
 \end{eqnarray}
which is local in the dual momentum $\omega'$.
Here, for convenience, we have defined the abbreviation
$$\hat \omega' \equiv e^{-2\gamma_E} \, \omega' \qquad 
\mbox{(and also $\hat\mu= e^{2\gamma_E} \, \mu$)} \,.$$
As a consequence the dual function $\rho_B^+(\omega')$
is renormalized multiplicatively,
\begin{eqnarray}
 \rho_B^+(\omega',\mu) &=&  
  e^{V} \left(\frac{\mu_0}{\hat \omega'} \right)^{-g} \, \rho_B^+(\omega',\mu_0)
  \nonumber \\[0.2em]
  &\equiv & U_{\omega'}(\mu,\mu_0) \, \rho_B^+(\omega',\mu_0) \,.
 \label{BmesonRGE1}
\end{eqnarray}
The RG functions $V=V(\mu,\mu_0)$ and $g=g(\mu,\mu_0)$ are expressed in terms of
the anomalous dimensions in the evolution kernel \cite{Lange:2003ff};
explicit expressions and a discussion of the composition rule of the RG elements,
\begin{align} U_{\omega'}(\mu_2,\mu_1) \, U_{\omega'}(\mu_1,\mu_0) &= U_{\omega'}(\mu_2,\mu_0) \,,
\label{comp}
\end{align} 
can be found in the appendix.

The transformation (\ref{trans}) thus diagonalizes the LN-kernel, which can also be seen by explicitly 
calculating the right-hand side of (\ref{eq:rge1}) for the identified 
\emph{continuous} set of eigenfunctions \footnote{It has recently been shown \cite{Braun:2014owa}
that (\ref{ev}) can be understood as the momentum-space representation of
the eigenvectors of one of the generators of collinear conformal transformations.
},
\begin{align}
 \phi_B(\omega) \ \to \ f_{\omega'}(\omega) & \equiv 
\sqrt{\frac{\omega}{\omega'}} \, J_1\left( 2 \, \sqrt{\frac{\omega}{\omega'}}\right) \,.
\label{ev}
\end{align}
In particular, using the 1-loop expression for the kernel, the non-local term in (\ref{eq:rge1}) yields
\begin{align}
 - \omega \, \int_0^\infty d\eta \, \Gamma^{(1)}(\omega,\eta) \, f_{\omega'}(\eta)
 &= - \Gamma^{(1)}_{\rm cusp} \, \ln \frac{\omega}{\hat\omega'} \, f_{\omega'}(\omega) \,,
\end{align}
which indeed combines with the local terms to the same RGE for $f_{\omega'}(\eta)$ as 
for $\rho_B^+(\omega')$ in (\ref{eq:rge2}), and the eigenvalues for the $f_{\omega'}(\omega)$
are 
\begin{align}
\gamma_{\omega'} & = -\left( \Gamma_{\rm cusp} \, \ln \frac{\mu}{\hat \omega'} + \gamma_+ \right) \,.
\label{gammawp}
\end{align}
The function $\rho_B^+(\omega')$ in dual momentum space thus plays a similar role as the
set of Gegenbauer coefficients for the LCDAs of light mesons 
\cite{Efremov:1979qk,Lepage:1979zb}, and the eigenfunctions $f_{\omega'}(\omega)$ are the analogue
of the Gegenbauer polynomials $C_n^{(3/2)}(2u-1)$ for the quark momentum fraction $u$ in a light meson. 
Notice however, that $\gamma_{\omega'}$ in (\ref{gammawp}) takes positive and negative values,
and therefore -- unlike for the case of the pion LCDA -- an asymptotic shape of the B-meson LCDA 
at (infinitely) large RG scales does not exist.
Still, as we will discuss below, perturbation theory implies model-independent constraints on 
the behaviour of $\rho_B^+(\omega')$.


\section{Positive Moments of $\phi_B^+$ \label{sec:3}}

Following \cite{Lee:2005gza}, we define positive moments of the LCDA $\phi_B^+(\omega,\mu)$ 
with a cut-off $\Lambda_{\rm UV}$ as
\begin{align}
 M_n(\Lambda_{\rm UV}) &:= \int\limits_0^{\Lambda_{\rm UV}} d\omega \, \omega^n \, \phi_B^+(\omega) \,.
\end{align}
For large $\Lambda_{\rm UV} \gg \Lambda_{\rm QCD}$, the moments are dominated by large values of 
$\omega$ in the integrand, and therefore can be estimated from a perturbative calculation based on
the \emph{partonic} result for the LCDA.
For the first two moments ($n=0,1$) one obtains the 1-loop expressions \cite{Lee:2005gza},
\begin{widetext}\begin{align}
 M_0(\Lambda_{\rm UV}) &= 1 + \frac{\alpha_s C_F}{4\pi} \left( -2 \,\ln^2 \frac{\Lambda_{\rm UV}}{\mu} + 
   2 \, \ln \frac{\Lambda_{\rm UV}}{\mu} - \frac{\pi^2}{12} \right)
    + \frac{16 \, \bar \Lambda}{3 \, \Lambda_{\rm UV}} \, 
    \frac{\alpha_s C_F}{4\pi} \left( \ln \frac{\Lambda_{\rm UV}}{\mu} -1 \right) + \ldots \,,
\nonumber \\
 M_1(\Lambda_{\rm UV}) &= \Lambda_{\rm UV} \, \frac{\alpha_s C_F}{4\pi} \left( -4 \, \ln \frac{\Lambda_{\rm UV}}{\mu} +6 \right)
 + \frac{4\bar\Lambda}{3} 
 \left[1+ \frac{\alpha_s C_F}{4\pi}\left( -2 \,\ln^2  \frac{\Lambda_{\rm UV}}{\mu} + 8 \, \ln  \frac{\Lambda_{\rm UV}}{\mu}
 - \frac74 - \frac{\pi^2}{12} \right) \right] + \ldots \,,
\label{pertmoments}
\end{align}
\end{widetext}
where the HQET-parameter $\bar\Lambda=m_B - m_b$ is defined in the pole-mass scheme.
Expressing the moments $M_n$ in terms of the dual function $\rho_B^+$, using (\ref{trans}),
we obtain
\begin{align}
 & M_n(\Lambda_{\rm UV})
 \cr &=
 \int\limits_0^\infty \frac{d\omega'}{\omega'} \int\limits_0^{\Lambda_{\rm UV}} 
 d\omega \, \omega^n \,  \sqrt{\frac{\omega}{\omega'}} 
  \, J_1\left(2 \sqrt{\frac{\omega}{\omega'}}\right) \rho_B^+(\omega') \,.
\end{align}
The $\omega$-integration can be performed explicitly, and for the first two
moments this yields
\begin{align}
 M_0(\Lambda_{\rm UV}) &= \Lambda_{\rm UV} \, \int\limits_0^\infty \frac{d\omega'}{\omega'} \,
 J_2\left( 2 \sqrt{\frac{\Lambda_{\rm UV}}{\omega'}} \right)  \rho_B^+(\omega') \,,
 \cr
 M_1(\Lambda_{\rm UV}) &= \frac{2 \Lambda_{\rm UV}}{3} \, M_0
 \cr & \quad {} -
 \frac{\Lambda_{\rm UV}^2}{3} \,\int\limits_0^\infty \frac{d\omega'}{\omega'} \, 
 J_4\left( 2 \sqrt{\frac{\Lambda_{\rm UV}}{\omega'}} \right)  \rho_B^+(\omega') \,.
 \cr &
  \end{align}

\subsection{Fixed-Order Matching}

Using properties of the Bessel functions summarized in the appendix,
we write the perturbative expansion for
the dual function as follows,
\begin{align}
& \rho_B^+(\omega')_{\rm pert.} =
  C_0(L) \, \frac{1}{\bar\Lambda} \, J_2\left(2\sqrt{\frac{2\bar\Lambda}{\omega'}}\right)
  \cr & \qquad 
  + \frac{4 \left( C_0(L) - C_1(L) \right)}{\bar\Lambda}
  \, J_4\left(2\sqrt{\frac{2\bar\Lambda}{\omega'}}\right)
  \cr & \qquad
  +\ldots \ ,
  \label{rhopert}
\end{align}
which reproduces the moments $M_0$ and $M_1$ up to further  
power corrections in $\bar \Lambda/\Lambda_{\rm UV}$.
At large values of $\omega' \gg \bar\Lambda$,
this can also be approximated by
\begin{align}
  \rho_B^+(\omega')_{\rm pert}
 &\quad \stackrel{\omega'\gg \bar\Lambda}{\simeq} \quad 
  C_0(L) \, \frac{1}{\omega'}
  - \frac23 \, C_1(L) \, \frac{\bar\Lambda}{(\omega')^2} + \ldots  
  \label{rhopertasym}
\end{align}
The coefficient functions at first order in the strong coupling
are obtained as
\begin{align}
 C_0(L) &= 1 + \frac{\alpha_s C_F}{4\pi}
 \left( - 2 L^2 + 2 L -2 - \frac{\pi^2}{12} \right) + {\cal O}(\alpha_s^2) \,,
 \cr 
 C_1(L) &= 
  1 + \frac{\alpha_s C_F}{4\pi}
 \left( - 2 L^2 + 2 L +\frac54 - \frac{\pi^2}{12} \right) + {\cal O}(\alpha_s^2)\,,
\label{Cn}
\end{align}
where the perturbative coefficients depend on logarithms $$L = \ln \frac{\mu}{\hat \omega'}\,.$$
We observe that at tree level ($C_0=C_1=1$) the expression in  (\ref{rhopert}) reduces
to the free parton model \cite{Kawamura:2001jm} as discussed in \cite{Bell:2013tfa},
\begin{align}
 \rho_B^+(\omega')_{\rm part.} &=
  \frac{1}{\bar\Lambda} \, J_2\left(2\sqrt{\frac{2\bar\Lambda}{\omega'}}\right) \,.
  \label{rhopar}
\end{align}

\subsection{(Local) RG Improvement}

\begin{figure}[tb]
 \centering
 \includegraphics[width=0.425\textwidth]{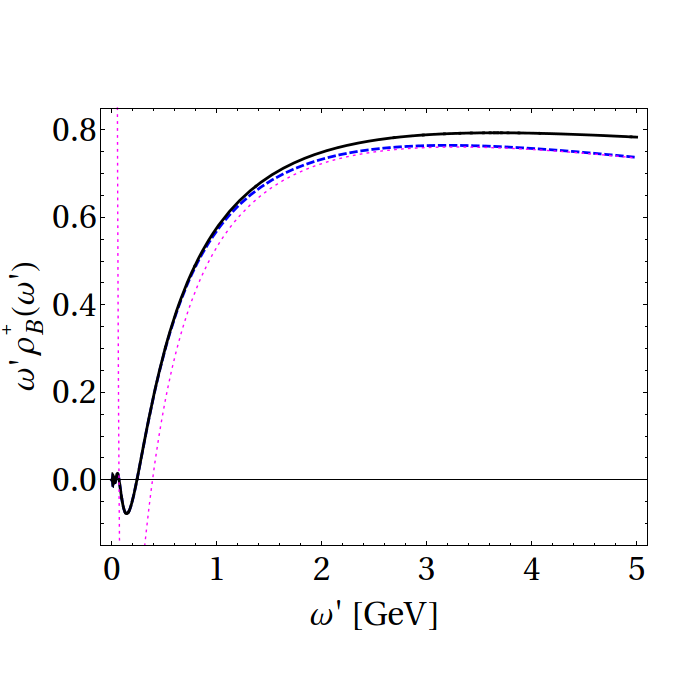} \vspace{-2em}
 \caption{\label{fig:rhocomp1} 
  Comparison of the perturbative expression for the dual function $\omega' \rho_B^+(\omega')$
  at $\mu=1$~GeV,
  with RG improvement (\ref{rhopertRG}, solid line) and without (\ref{rhopert}, dashed line).
  The dotted line shows the expansion for large $\omega'$ in (\ref{rhopertasym}).
  The HQET parameter $\bar\Lambda$ has been set to $465$~MeV.}
\end{figure}

\begin{figure}[tb]
 \centering
 \includegraphics[width=0.425\textwidth]{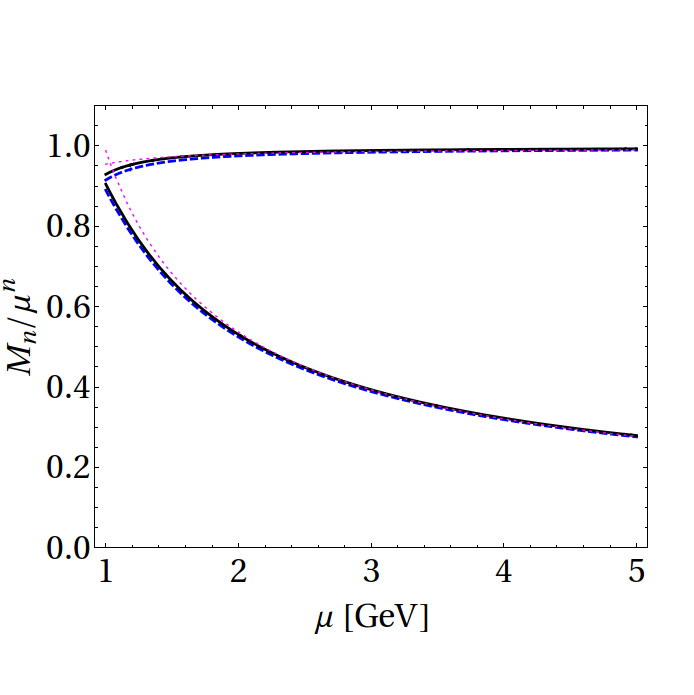} \vspace{-2em}
 \caption{\label{fig:Mncomp1} 
 Comparison of regularized moments $M_{0,1}$ as a function of $\mu$
with the UV cutoff set to $\Lambda_{\rm UV} = e^{\gamma_E} \, \mu$,
 computed from the 
  perturbative expression for the dual function 
  with RG improvement (\ref{rhopertRG}, solid lines) and without (\ref{rhopert}, dashed lines).
  The dotted line shows the result from the direct computation (\ref{pertmoments}).
  Here $M_0$ evolves towards 1, while $M_1/\mu$ drops proportional to $\alpha_s$. 
  The HQET parameter $\bar\Lambda$ has been set to $465$~MeV.}
\end{figure}

The logarithms $L$ in (\ref{Cn}) become large for values of $\omega'$
much smaller or larger than $\mu$.
As the coefficients $C_n$ in (\ref{Cn}) fulfill the same 1-loop RGE as the dual function $\rho_B^+$,
\begin{align}
 &\frac{d}{d\ln\mu} \, C_i
 =  \frac{\alpha_s C_F}{4\pi} \left( - 4 L + 2 \right) C_i + {\cal O}(\alpha_s^2)
 \cr 
 &= - \frac{\alpha_s C_F}{4\pi} \left(  \Gamma_{\rm cusp}^{(1)} \, \ln \frac{\mu}{\hat \omega'}
    + \gamma_+^{(1)} \right) C_i + {\cal O}(\alpha_s^2) \,,
\end{align}
we may resum perturbative logarithms into the RG function $U_{\omega'}$ as long as $\omega'$ is 
sufficiently large. To this end, we define
\begin{align}
\mu_{\omega'} = \mu_{\omega'}(\mu)& := \sqrt{ (k \, \hat\omega')^2 + \mu^2 } \,, 
 \label{muw}
\end{align}
with a numerical parameter $k$ with default value 1, such that $\mu_{\omega'} \sim \hat\omega'$
for large values of $\omega'$, and $\mu_{\omega'} \sim \mu$ at small values of $\omega'$.
With this we obtain an RG-improved expression for $\rho_B^+(\omega')_{\rm pert}$,
\begin{align}
 \rho_B^+(\omega',\mu)_{\rm RG} &= U_{\omega'}(\mu,\mu_{\omega'}) \, \rho_B^+(\omega',\mu_{\omega'})_{\rm pert} \,,
 \label{rhopertRG}
\end{align}
which is valid for $\omega'\gtrsim \mu$. 
(Notice that the implicit $\mu$-dependence from the auxiliary scale $\mu_{\omega'}$
cancels between the two factors in (\ref{rhopertRG}) up to higher-order corrections
which will be numerically checked by varying the parameter $k$.)
The RG-improved form (\ref{rhopertRG}) is
compared to the result from fixed-order perturbation theory (FOPT) 
in (\ref{rhopert}) in Fig.~\ref{fig:rhocomp1}.
As we discuss below, our new idea of local RG improvement is
essential for the understanding of the asymptotic behaviour of $\rho_B^+$ for $\omega' \to \infty$.
However, this resummation will only slightly modify the positive moments $M_{0,1}$ compared to
the FOPT expressions in (\ref{pertmoments}), as long as $\mu$ and $\Lambda_{\rm UV}$
are sufficiently large and of similar size. This is illustrated in Fig.~\ref{fig:Mncomp1}.

For a given (high) scale $\mu \gg \bar\Lambda$ and large values of $\hat\omega'\gg \mu$, 
the dual function is thus completely determined perturbatively --- independent of any hadronic model --- 
with
\begin{align}
 \rho_B^+(\omega',\mu) &\stackrel{\hat\omega'\gg \mu}{\longrightarrow}  \ e^{V(\mu,\hat\omega')} 
 \rho_B^+(\omega',\mu=\hat\omega')_{\rm pert.}  \,.
 \label{pertbehave}
\end{align}

\section{Construction of $\rho_B^+(\omega')$ \label{sec:4}}

Our aim is now to find a systematic parametrization 
for the dual function $\rho_B^+(\omega',\mu)$ which
interpolates between some low-energy model (valid
at small renormalization scales and small values of $\omega'$)
and the perturbative behaviour in (\ref{pertbehave}), with
the following features:
\begin{itemize}
 \item Explicit implementation of the RG evolution as discussed above.
 \item Correct behaviour at large values of $\omega'$, such that 
   the constraints on positive moments of $\phi_B^+(\omega,\mu$)
   in HQET, as discussed above, are fulfilled.
\end{itemize}
Starting from a model function $\rho^{\rm model}(\omega')$, which 
is supposed to  have a Taylor expansion in $1/\omega'$ at
large values of $\omega'$ 
and to  give a good description of the low-$\omega'$ region,
we then propose the following ansatz
\begin{align}
&\rho_B^+(\omega',\mu) :=
\cr &  \phantom{+} U_{\omega'}\left(\mu,\mu_{\omega'}(\mu_0)\right)  \left[ 
\rho^{\rm model}(\omega')  - \sum_{n=0}^N D_n^{\rm model} \, p_n(\omega')\right]
\cr 
& {} + U_{\omega'}\left(\mu,\mu_{\omega'}(\mu)\right) \sum_{n=0}^N 
  D_n^{\rm pert}\left(\ln \frac{\mu_{\omega'}(\mu)}{\hat\omega'},\mu_{\omega'}(\mu)\right)   p_n(\omega') 
   \,.
   \cr &
 \label{ansatz}
\end{align}
In the first line we start with a given model $\rho^{\rm model}$ 
for the dual function and subtract a number of terms,
with $\{p_n\}$ representing 
a set of appropriate functions which reduce to a power-law behaviour $1/(\omega')^{n+1}$ 
for large values of $\omega'$, 
and vanish quickly at small $\omega'$.
The term in square brackets then reduces to $\rho^{\rm model}$
for small values of $\omega'$ but now decreases as $1/(\omega')^{N+1}$
at large values of $\omega'$.
The evolution factor in front is chosen to refer to 
a hadronic reference scale $\mu_0$ assigned to the input model \footnote{%
If the hadronic model under consideration gives an explicit reference
to the renormalization scale $\mu_0$, one would have to replace $\mu_0 \to \mu_{\omega'}(\mu_0)$ in 
$\rho^{\rm model}(\omega',\mu_0)$.}.
The term in the second line uses the same set of basis functions
to reproduce the RG-improved perturbative result in (\ref{rhopertRG}).

In the following analysis, we take the first two coefficients in that sum into account ($N=1$).
The coefficients $D_n$ can then be matched by expanding the corresponding 
perturbative or model functions for large values of $\omega'$, which
will be done below.
For the set of functions $p_n$ we choose
\begin{align}
p_n(\omega') &\equiv \frac{\Omega^n}{(\omega'+\Omega)^{n+1}} \, e^{-\Omega/\omega'} \,,
\label{pndef}
\end{align}
so that the modifications from the perturbative matching are exponentially suppressed
at small values of $\omega'$, where the original model is supposed to give
a reasonable functional description. 
The auxiliary parameter $\Omega$ sets the scale where the transition between the
perturbative and non-perturbative regime occurs.

\subsection{Models}

The coefficients $D_n$ can easily be 
extracted
by comparing the Taylor expansion in $1/\omega'$.
For instance, for the exponential model
\begin{align}
\rho^{\rm model-1}(\omega') &= \frac{1}{\omega'} \, e^{-\omega_0/\omega'} \,,
\label{expmodel}
\end{align}
one obtains 
\begin{align}
 D_0^{\rm model-1} &= 1 \,, \ \  D_1^{\rm model-1} = {2} - \frac{\omega_0}{\Omega} \,.
\end{align}
and for the free parton model \cite{Kawamura:2001jm}, as discussed in \cite{Bell:2013tfa},
\begin{align}
\rho^{\rm model-2}(\omega') &= \frac{1}{\bar\Lambda} \, J_2\left(2 \sqrt{\frac{2\bar\Lambda}{\omega'}} \right)  \,,
\end{align} 
one gets
\begin{align}
 D_0^{\rm model-2} &= 1 \,, \ \  D_1^{\rm model-2} = {2} - \frac{2\bar\Lambda}{3\Omega} \,.
\end{align}
As a third illustrative model, we consider the tree-level estimate of a QCD sum-rule
analysis in \cite{Braun:2003wx} where $\phi_B^+(\omega)=3/4/\omega_0^3 \,\theta(2 \omega_0 - \omega)
\, 
\omega \, (2 \omega_0 - \omega)$ which corresponds to
\begin{align}
 \rho^{\rm model-3}(\omega') &= \frac{3}{\omega_0}
  \, \sqrt{\frac{\omega'}{2\omega_0}} \, 
  J_3\left(2 \, \sqrt{\frac{2\omega_0}{\omega'}} \right) \,,
\end{align}
and
\begin{align}
 D_0^{\rm model-3} &= 1 \,, \ \  
 D_1^{\rm model-3} = 2 - \frac{\omega_0}{2\Omega} \,.
\end{align}

\subsection{Matching with OPE \label{sec:4A}}

The ansatz (\ref{ansatz}) reduces straight-forwardly to an expression in FOPT by setting $\mu_{\omega'}(\mu)=\mu_{\omega'}(\mu_0)=\mu$.
The matching coefficients $D_n^{\rm pert}$ are then easily obtained by equating
the resulting large-$\omega'$ expansion with (\ref{rhopertasym}).
For the first two coefficients, 
we obtain
\begin{align}
 D_0^{\rm pert}\left(\ln \frac{\mu_{\omega'}}{\hat\omega'},\mu_{\omega'}\right) &=  
   C_0\left(\ln\frac{\mu_{\omega'}}{\hat\omega'},\mu_{\omega'}\right) \,, \cr 
 D_1^{\rm pert}\left(\ln\frac{\mu_{\omega'}}{\hat\omega'},\mu_{\omega'}\right) &= 
   2 C_0\left(\ln \frac{\mu_{\omega'}}{\hat\omega'},\mu_{\omega'}\right) 
   \cr & \quad - 
   \frac{2\bar\Lambda}{3\Omega} \, C_1\left(\ln\frac{\mu_{\omega'}}{\hat\omega'},\mu_{\omega'}\right) \,.
\end{align}
Notice that, at this stage, the parameter $\Omega$ is arbitrary.
However, as we will see in the numerical analysis, 
the dependence of the $\rho_B^+(\omega,\mu)$
on the value of $\Omega$ is not very pronounced.
For concreteness, we will use a default value of
$\Omega = e^{\gamma_E} \mu_0$. 


\section{Logarithmic Moments}

\label{sec:4a}
 
The logarithmic moments of the dual function can be defined as 
\begin{align}
L_k(\mu) \equiv
 \int\limits_0^\infty \frac{d\omega'}{\omega'} \,  \ln^k \left(\frac{\hat\omega'}{\mu} \right) \rho_B^+(\omega',\mu)\,.
\label{Lmom}
 \end{align}
As emphasized in \cite{Bell:2013tfa} the first three logarithmic moments ($k=0,1,2$) of $\phi_B^+(\omega)$ are identical 
to those of $\rho_B^+(\omega')$.
They represent the hadronic input parameters appearing in factorization theorems for exclusive $B$-meson decays
to first non-trivial order in the strong coupling constant.

\begin{itemize}
 \item Contributions from $\hat\omega'\geq \mu$ are completely determined perturbatively
   via (\ref{pertbehave}),
 \begin{align}
& L_k^+(\mu) \equiv
 \cr 
 & \quad \int\limits_\mu^\infty \frac{d\hat\omega'}{\hat\omega'} \,  
 \ln^k \left(\frac{\hat\omega'}{\mu} \right) 
  e^{V(\mu,\hat\omega')} 
 \rho_B^+(\omega',\mu=\hat\omega')_{\rm pert.}
 \,.
\label{Lmomplus}
 \end{align}  
 This will be illustrated and confirmed numerically 
 in Sec.~\ref{sec:numerics}.
   
 \item For the contributions from $\hat\omega'\leq \mu$,
 substituting $z=-\ln \frac{\hat\omega'}{\mu}$,
 \begin{align}
L_k^-(\mu) \equiv
 \int\limits_0^{\infty} dz \, (-z)^k \, 
 \rho_B^+(\hat\mu \, e^{-z},\mu)
 \,,
\label{Lmomminus}
 \end{align}  
 we may expand the 
   function $\rho_B^+(\omega',\mu)$ in terms
   of Laguerre polynomials ${\rm L}_n(z)$,
   \begin{align}
    \rho_B^+(\hat\mu \, e^{-z},\mu)
    &:= \sum_{n=0}^\infty a_n(\mu) \, e^{-z} \, {\rm L}_n(z) \,.
   \end{align}
  For a given model, the coefficients $a_n(\mu)$
  can be obtained from the orthogonality relation
  (\ref{ortho_Laguerre}). The logarithmic moments
  simply follow as
 \begin{align}
   L_0^-(\mu) &= a_0(\mu) \,, \cr 
   L_1^-(\mu) &= a_1(\mu)-a_0(\mu) \,, \cr 
   L_2^-(\mu) &= 2 a_2(\mu) - 4 a_1(\mu) + 2 a_0(\mu) \,,
   \cr & \mbox{etc.}
  \end{align}
\end{itemize}

In principle the first few $L_k^-$ -- and hence the first few $a_k$ --
can be determined from precision analyses of radiative leptonic
$B$-meson decays (see \cite{Beneke:2011nf,Braun:2012kp} for recent
discussions). On the theory side the task is difficult.  When more
information on the  perturbative analysis of the moments $M_n$
becomes available, it only affects the precision of our knowledge of $L_k^+$, but does not
spill over into the non-perturbative contribution. It may be
interesting to see if non-perturbative methods like sum rules and
lattice QCD are able to shed more light on the Laguerre coefficients
in $L_k^-$, once the dual function $\rho_B^+$ is used in lieu of $\phi_B$.

\subsection{Large-Scale Behaviour}

\begin{figure}[htb]
 \centering
 \includegraphics[width=0.425\textwidth]{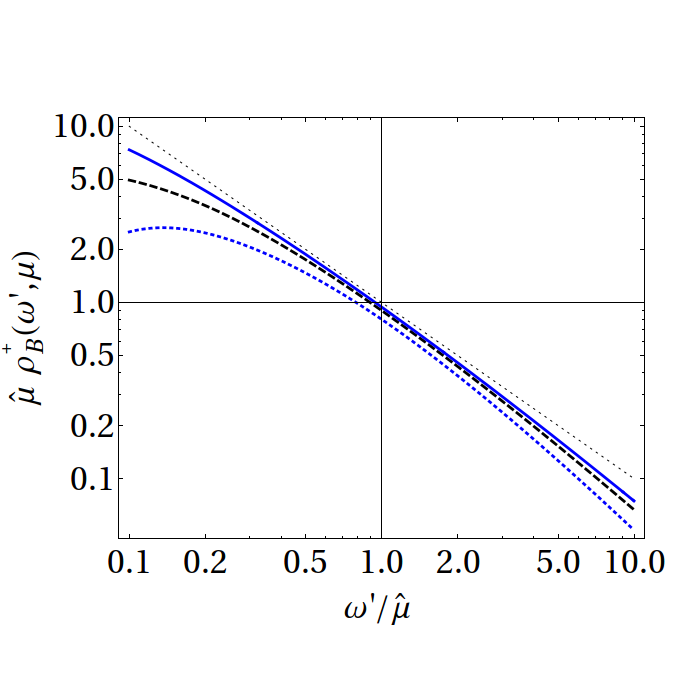} \vspace{-2em}
 \caption{\label{fig:loglog} 
  Double logarithmic plot for the behaviour of $\hat \mu \, \rho_B^+(\omega',\mu)$ 
  for model-1 at different renormalization scales, $\mu=1$~GeV (thick dotted line),  $\mu=3$~GeV (dashed line),
  $\mu=10$~GeV (solid line), compared to the naive asymptotic $\hat \mu/\omega'$ behaviour 
  at large values of $\omega'$ (thin dotted line, which is the only one that truly goes through the point $(1,1)$).}
\end{figure}

As we have seen, at tree-level in FOPT
the dual function $\rho_B^+(\omega')$ behaves as $\sim 1/\omega'$ 
for large values of $\omega'$, see (\ref{rhopertasym}).
This behaviour is softened by (global)
evolution towards higher scales, i.e.\ 
$$g=g(\mu,\mu_0) > 0 \qquad \mbox{for $\mu >\mu_0$}$$ 
in (\ref{BmesonRGE1}), which induces an additional factor
$(\omega')^g$. 
Therefore it appears  as if for
sufficiently large values of $g$ the $\omega'$-integrals which, for
instance, determine the transformation back to $\omega$-space in
(\ref{trans}) or the logarithmic moments in (\ref{Lmom}), would no
longer converge.  
(In other words it seems as if $\rho_B$ undergoes
a qualitativ change by evolving to sufficiently large scales $\mu \gg
\mu_0$ such that $g(\mu,\mu_0) \ge 1$.)
The local RG improvement  as
it is implemented in (\ref{rhopertRG}) reveals, however, that the perturbative resummation
of logarithms $\ln \omega'/\mu$ from the region $\omega' \gg \mu$ always yields converging
results, since at large (but fixed) values of $\mu$ one rather has
$$
\rho_B^+(\omega',\mu) \sim (\omega')^{-1-g(\mu_{\omega'},\mu)}
$$
and $g(\mu_{\omega'},\mu) > 0$ for $\hat\omega' \gg \mu$.
Therefore, in that asymptotic region, the function $\rho_B^+(\omega')$ 
always decreases \emph{faster} than $1/\omega'$, which can also be
seen from Fig.~\ref{fig:loglog} where we compare the behaviour
of $\rho_B^+(\omega')$ at different renormalization scales.

For phenomenological applications in exclusive
$B$-meson decays one always has $g(\mu,\mu_0)<1$, but we may still formally 
consider the case $g\geq 1$ for curiosity.
From the discussion in the previous paragraph, we conclude that 
the solution of the RGE for the dual function $\rho_B^+(\omega',\mu)$ 
makes sense for arbitrary values of $\mu$ and $\mu_0$ (provided they are sufficiently
larger than $\Lambda_{\rm QCD}$). 
In contrast, the derivation of the RGE solutions for 
the original LCDA in (\ref{Bdef}) --
as discussed in \cite{Lee:2005gza,Kawamura:2001jm} -- is formally restricted to values $0<g(\mu,\mu_0)<1$.
So we repeat for clarity's sake that the logarithmic moments $L_k$ exist at {\em all}
scales $\mu\gg \Lambda_{\rm QCD}$, and that perceived thresholds are mathematical artifacts.

 
\section{Numerical Examples \label{sec:5}}

\label{sec:numerics}

\subsection{Preliminaries}

\paragraph*{Mass-renormalization and $\bar\Lambda$:}
For the numerical analysis, we take the $b$-quark mass
from a determination in a different mass scheme,
the shape-function (SF) scheme \cite{Bosch:2004bt},
which is related to the pole-mass scheme via
\begin{align}
 \bar\Lambda &= \bar\Lambda_{\rm SF}(\mu_f,\mu) - \mu_f \, \frac{\alpha_s(\mu) \, C_F}{\pi} 
 \left( 1-2\, \ln \frac{\mu_f}{\mu} \right) \,,
\end{align}
with a fixed reference scale $\mu_f \sim \mu$.
The HFAG~2013 update 
\cite{Amhis:2012bh} quotes 
$$
  \bar\Lambda_{\rm SF}(\mu_\ast,\mu_\ast) = (0.691 \pm 0.025)~{\rm GeV} \,,
$$for a common value $\mu=\mu_f=\mu_\ast=1.5$~GeV.
For the numerical discussion in the pole-mass scheme \footnote{In this work we do not
entertain the idea to change the mass scheme in the perturbative matching procedure itself.
Although a renormalon-free scheme like the one proposed in
\cite{Lee:2005gza} would improve the perturbative convergence for the 
regularized moments $M_n$, we do not expect a significant effect for the logarithmic
moments relevant for phenomenological applications since, as we will show below, the 
contributions from the perturbative regime are subdominant there. }, this corresponds to using
$$
 \bar\Lambda = (0.465 \pm 0.025)~{\rm GeV} \,.
$$

\paragraph*{Hadronic reference scale:}
As our default choice for a hadronic reference scale,
we use 
$$
 \mu_0 =1.0~{\rm GeV} \,.
$$
The dimensionless parameter $k$ in (\ref{muw}) is taken at
a default value $k=1$ and varied between $k=1/2$ and $k=2$.
As already mentioned above, 
our default choice for the scale-parameter $\Omega$
in the functions $p_n(\omega')$ is set to
$$
  \Omega = e^{ \gamma_E}\, \mu_0 \simeq 1.78~{\rm GeV} \,. 
$$
Again this value will be varied within a factor of $2$ to study the sensitivity
of our parametrization with respect to this parameter. \\

\paragraph*{Running coupling constant:}
For $\alpha_s(\mu)$ we take the 3-loop formula (\ref{as3}) with a fixed number of
flavours $n_f=4$, and $\Lambda_{\rm QCD}^{(4)} \simeq 299$~{\rm MeV},
which corresponds to
$$
 \alpha_s(1~{\rm GeV}) \simeq 0.466\,,\qquad \alpha_s(5~{\rm GeV}) \simeq 0.214 \,.
$$
We also take into account the 3-loop $\beta$-function in the evaluation
of the RG function $g$ (see appendix). \\

\paragraph*{Model 1:}
The parameter $\omega_0$ 
in (\ref{expmodel}) at $\mu=1.0$~GeV is set to $438$~{\rm MeV},
where we have used the value advocated in \cite{Lee:2005gza}. \\

\paragraph*{Model 2:}
The free parton model does not involve additional
hadronic parameters, except for the HQET parameter 
$\bar\Lambda$, together with the low-energy
reference scale $\mu_0$, which have already been fixed above.\\

\paragraph*{Model 3:}
The tree-level sum rule estimate in \cite{Braun:2003wx}
contains a parameter $\omega_0=1.0$~GeV at $\mu_0=1.0$~GeV. \\

\subsection{Illustrations}
\paragraph*{$\rho_B^+$ at large values of $\omega'$:}
In Fig.~\ref{fig:mod1_high}, we present the result for the product
$\omega' \rho_B^+(\omega,\mu)$ at large values $\omega'>\mu_0$ for two
choices of renormalization scale, $\mu_0=1$~GeV and $\mu=5$~GeV. 
We observe
that the inclusion of the radiative corrections
shows a significant effect compared to the original (tree-level) functions
$\rho^{\rm model}(\omega')$, while the differences among
the different models is irrelevant at large values of $\omega'$. \\

\paragraph*{$\rho_B^+$ at small values of $\omega'$:} 
In Fig.~\ref{fig:mod1_low}, we present the results for $\rho_B^+(\omega,\mu)$
at small values $\omega'<\mu_0$ for two choices of renormalization scale, $\mu_0$ 
and $\mu=5$~GeV. We observe that for our default value of the parameter $\Omega$
the original model is reproduced extremely well, and the variation of this parameter
has only a minor effect at intermediate values of $\omega'$.
\\

\paragraph*{The LCDA $\phi_B^+(\omega)$:} 
From the QCD-improved dual function $\rho_B^+(\omega',\mu)$ in (\ref{ansatz}) for a given
input model, we can easily compute the corresponding LCDA $\phi_B^+(\omega,\mu)$
via (\ref{trans}) by numerical integration. 
We have compared the
original (tree-level) models of $\phi_B^+(\omega,\mu_0)$ and their QCD-improved versions
for the 3 benchmark models. For all three models we recovered the feature 
of a (negative) ``radiative tail''
at large values of $\omega$ \cite{Lee:2005gza}, while the
behaviour at small values of $\omega$ is practically unaffected. \\

\paragraph*{Regularized moments of $\phi_B^+$:} 
In Table~\ref{tab:opemom} we list the first two (regularized) moments
$M_0$ and $M_1$ as obtained from different models and different renormalization scales,
and compare them to the perturbative result obtained from the local RG-improved
formula (\ref{rhopertRG}). Here, the value for the UV cutoff is chosen slightly
larger than the renormalization scale, $\Lambda_{\rm UV} := e^{\gamma_E} \, \mu$,
in order to assure that the result is sufficiently dominated by the radiative
tail in the corresponding LCDA $\phi_B^+(\omega)$.
We observe that the zeroth moment is reproduced rather 
accurately by the different models; the first moments differ more, around 10\%\ 
at $\mu=10$~GeV. These 
differences are easily explained by the fact that those higher-order terms 
in $M_1$ which are not fixed by the matching procedure are of the order
$1/\alpha_s \cdot \Omega^2/\Lambda_{\rm UV}^2$ relative to $M_1$. \\

\begin{table}[t]
\renewcommand{\arraystretch}{1.2}
\centering
 \begin{tabular}{ c
 |>{\centering\let\newline\\\arraybackslash\hspace{0pt}}m{3.8em}
 |>{\centering\let\newline\\\arraybackslash\hspace{0pt}}m{3.8em}
 |>{\centering\let\newline\\\arraybackslash\hspace{0pt}}m{3.8em}  
 |>{\centering\let\newline\\\arraybackslash\hspace{0pt}}m{3.8em}
 |>{\centering\let\newline\\\arraybackslash\hspace{0pt}}m{3.8em}
 }
 \hline \hline 
  & $\mu$ & model-1 & model-2 & model-3 & 
  pert.\ (RG)
  \\
  \hline \hline
      & 3 GeV & 0.996 & 0.995 & 0.998 & 0.988
  \\    
  $M_0$ & 6 GeV & 1.032 & 1.024 & 1.036 & 0.993
  \\
      &10 GeV & 1.011 & 1.004 & 1.014 & 0.995
  \\ \hline 
      &3 GeV & 0.454 & 0.383 & 0.494 & 0.393
  \\    
  $M_1/\mu$ & 6 GeV & 0.329 & 0.287 & 0.351 & 0.250
  \\
      &10 GeV & 0.207 & 0.183 & 0.219 & 0.188
  \\ \hline \hline
 \end{tabular}
  \caption{\label{tab:opemom} 
Comparison between the regularized
moments $M_0$ and $M_1$ from the QCD-improved model functions for $\phi_B^+$ 
and the RG-improved perturbative results from  (\ref{rhopert},\ref{rhopertRG})
at different values of $\mu$ with $\Lambda_{\rm UV}= e^{\gamma_E}\,\mu$.
}
 \end{table}

\paragraph*{Logarithmic Moments:} 

\begin{table}[b]
\renewcommand{\arraystretch}{1.2}
\centering
 \begin{tabular}{ l | >{\centering\let\newline\\\arraybackslash\hspace{0pt}}m{6em}|>{\centering\let\newline\\\arraybackslash\hspace{0pt}}m{6em}|>{\centering\let\newline\\\arraybackslash\hspace{0pt}}m{6em} }
 \hline \hline 
 $L_n$ & \centering total & \centering from $\hat\omega'<\mu$ & from $\hat\omega'\geq \mu$
  \\
 \hline \hline  
  $L_0$ (model 1) & \centering 1.67 & \centering 1.58 &  0.086 
  \\
  $L_0$ (model 2) &\centering 1.65 &\centering  1.57 &  0.086
  \\
  $L_0$ (model 3) &\centering 1.21 & \centering 1.12 &  0.086
  \\ \hline 
  $L_1$ (model 1) &  \centering -3.85 &\centering -3.93 & 0.074
  \\
  $L_1$ (model 2) &  \centering -3.46 &\centering -3.54 & 0.074
  \\
  $L_1$ (model 3) & \centering -2.19 &\centering -2.27 & 0.074 
  \\ \hline 
  $L_2$ (model 1) &\centering 11.6  &\centering 11.4  &  0.121
  \\
  $L_2$ (model 2) & \centering 9.03 &\centering 8.91 & 0.121
  \\
  $L_2$ (model 3) &\centering 5.44 &\centering 5.32 & 0.121
  \\ \hline  
 \end{tabular} 
\\ $\phantom{a}$ \\[2mm]
\begin{tabular}{ l | >{\centering\let\newline\\\arraybackslash\hspace{0pt}}m{6em}|>{\centering\let\newline\\\arraybackslash\hspace{0pt}}m{6em}|>{\centering\let\newline\\\arraybackslash\hspace{0pt}}m{6em} }
 \hline  \hline 
 $L_n$ & \centering total (RG)  &  FOPT & input (tree)
 \\
 \hline \hline 
  $L_0$ (model 1) & 2.19  &2.12  & 2.28
  \\
  $L_0$ (model 2) & 2.05  & 1.98 & 2.15
  \\
  $L_0$ (model 3) & 1.40 &  1.34  & 1.50
  \\ \hline 
  $L_1$ (model 1) & -3.31  & -3.45 &  -3.20
  \\
  $L_1$ (model 2) & -2.41   & -2.56 & -2.31
  \\
  $L_1$ (model 3) & -1.31   &  -1.46 &  -1.21
  \\ \hline 
  $L_2$ (model 1) & 7.88 & 7.19 &  8.25
  \\
  $L_2$ (model 2) & 4.25   & 3.56 & 4.62
  \\
  $L_2$ (model 3) & 2.48  & 1.80 & 2.85
  \\ \hline \hline 
 \end{tabular}
\caption{\label{tab:logmom} 
Logarithmic moments $L_n$  from different models (values in
GeV$^{-1}$). The theoretical uncertainties are dominated by the hadronic input model. 
{\em Above:} Comparison between the contributions from low and high $\omega'$ regions
  at $\mu=3$~GeV. 
{\em Below:}  Effects of the perturbative and local RG improvement at $\mu_0=1$~GeV.
}
\end{table}

In Table~\ref{tab:logmom}, we compare the numerical results for 
the first three logarithmic moments $L_{0,1,2}$ following from the
three different benchmark models.
In the upper part, we consider an intermediate scale
$\mu=3$~GeV and separate the contributions to the integral from regions where $\hat\omega'$ is smaller or larger than $\mu$.
We observe that the former -- by construction -- very much depend on the specific hadronic input model.
The contributions to the moments 
from large values, $\hat\omega'>\mu$, on the other hand,
are completely determined by perturbative matching and RG evolution 
and therefore independent of the
hadronic input model, in line with the discussion around
(\ref{pertbehave}).
In the lower half of the table, we illustrate the effects of the perturbative constraints
by comparing the moments originating from the hadronic input function with its modifications
from FOPT alone and its (locally) RG-improved version, at the hadronic input scale $\mu_0=1$~GeV.
Again, we observe that the logarithmic moments before and after the perturbative improvement
are highly correlated and do not differ very much.\\


\section{Summary}

The dual function $\rho_B^+(\omega')$ of the heavy B-meson, which
plays a similar role to the familiar set of Gegenbauer coefficients
for light-meson LCDAs, has been the subject of this paper. To
recapitulate and summarize we have highlighted that the transformation
between the LCDAs in momentum space, $\phi_B^+(\omega)$, and dual
momentum space, $\rho_B^+(\omega')$, consists of eigenfunctions of the
RG evolution kernel for $\phi_B^+(\omega)$. 
The dual function renormalizes multiplicatively, unlike the LCDA
$\phi_B^+$, which in particular implies that the non-perturbative low-$\omega'$ region
does not mix with the perturbative domain, $\omega' \gtrsim \mu$.

We have demonstrated that the dual function in the perturbative
domain is calculable from the OPE results of a finite-moment analysis
of $\phi_B^+$ \cite{Lee:2005gza}, and determined its values in the
region $\omega' \sim \mu$. By resumming perturbative logarithms we
have shown that $\rho_B$ vanishes faster than $1/\omega'$ for $\omega'
\gg \mu$, such that its first inverse moment, $\lambda_B^{-1}(\mu)$,
converges for $\omega'\to\infty$ at any perturbative scale $\mu$. This
result distinguishes the analysis of $\lambda_B$ and related
quantities in dual momentum space from the corresponding one using the
standard LCDA $\phi_B^+$, which appears to suffer from artificial thresholds in
its RG evolution when the RG function $g(\mu,\mu_0)$ assumes positive integer
values.

The low-$\omega'$ regime is not accessible via perturbation theory. It
must be determined by other means, but can be modelled in the
meantime. We have developed a method of combining a model ansatz in
the non-perturbative regime with the perturbative results that
respects the moment constraints and smoothly connects both
domains. This was achieved by correcting the first few terms in the
large-$\omega'$ behaviour of the model, which is shown in
Eq.~(\ref{ansatz}). The keen-eyed reader might inquire
whether taking ever more terms into account in this manner will
ultimately determine the dual function and thus render the
modelled part superfluous. We found only poor or no convergence with
such an approach,  depending on the choice of basis functions $p_n$,
which points towards an unrelatedness between
$\lambda_B$ and other HQET parameters like $\bar\Lambda$ within
perturbative methods.

We have illustrated our results using three different models, and also
performed a numerical analysis for the phenomenologically relevant
logarithmic moments $L_k$. Following the gist of our discussion thus
far, we split $L_k=L_k^++L_k^-$ between the regions $\hat\omega' >
\mu$ and $\hat\omega' < \mu$. While the contributions $L_k^+$ are
model-independent (and small), the hadronic model dominates the
moments. In principle the $L_k^-$ can be determined from precision
analyses of radiative leptonic $B$-meson decays,
but this only determines the first few terms in an expansion 
of the dual function in terms of Laguerre polynomials in 
the variable $z=\ln \mu/\hat\omega'$ for $\hat\omega'\leq \mu$.

The perturbative analysis of the $B$-meson LCDA is restricted
to 1-loop accuracy so far. From the theoretical perspective, it
would be interesting to see to what extent the picture that
emerged from our analyses continues to be valid when implementing 
2-loop corrections to the evolution kernel and the 
perturbative moment constraints, together with higher 
power corrections in HQET.

\begin{acknowledgments}

TF  and BOL acknowledge support by the Deutsche Forschungsgemeinschaft (DFG)
within the Research Unit {\sc FOR~1873} ({\it Quark Flavour Physics and Effective Field Theories}).
YMW  is supported
by the  {\it DFG-Sonder\-forschungs\-bereich/Transregio 9 ``Computergest\"{u}tzte Theoretische Teilchenphysik''}.
We are particularly grateful to Guido Bell who continually accompanied the project and 
contributed many valuable discussions and comments and critical reading of the manuscript.
TF also would like to thank Thomas Mannel for helpful discussions.

\end{acknowledgments}


\begin{widetext}
\begin{figure*}
 \centering
 \includegraphics[width=0.425\textwidth]{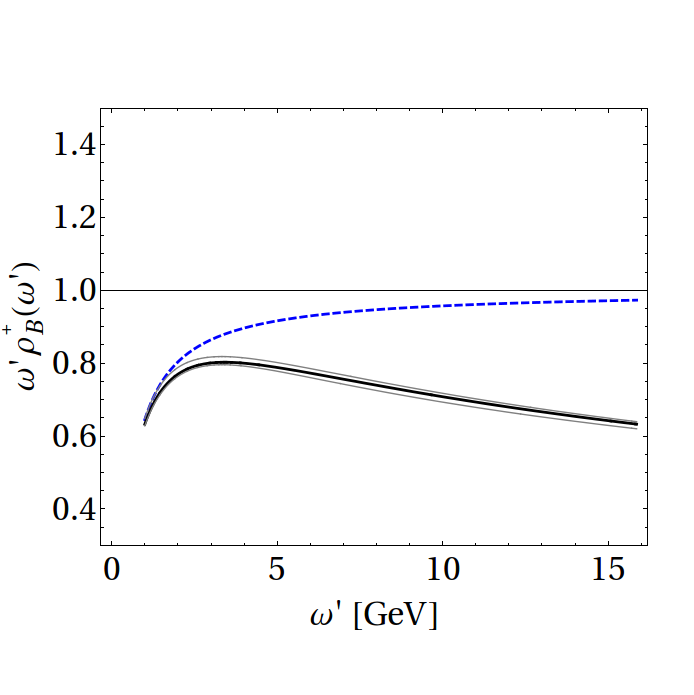} \quad 
 \includegraphics[width=0.425\textwidth]{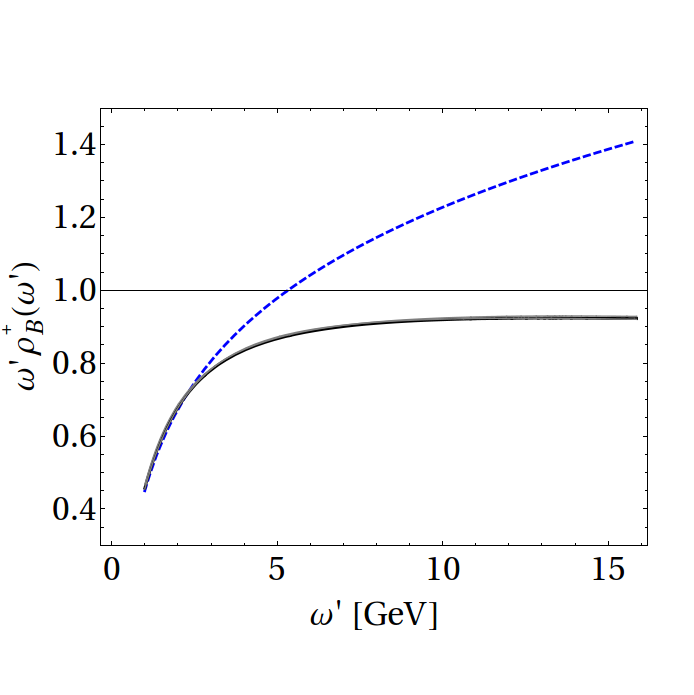} 
 \caption{\label{fig:mod1_high} Behaviour of the dual function
   $\rho_B^+(\omega')$ (multiplied with a factor $\omega'$) at large
   values of $\omega'$. 
   We show the result for model-1; for models-2,3 the result
   looks almost identical.
   Left: For $\mu=\mu_0$. Right: For $\mu=5$~GeV. The solid lines represent
   the result of the RG-improved parametrization
   (\ref{ansatz}). The grey lines indicate the variation
   of the parameter $k=1/2,2$ in the defintion of the auxiliary scale
   $\mu_{\omega'}$. The thick (blue) dashed lines refer to the original model
   function $\rho^{\rm model-1}(\omega')$.
   (Notice that in the right plot,
   the model function is [globally] evolved from $\mu_0 \to 5$~GeV.)}
\end{figure*}

\begin{figure*}
 \centering
 \includegraphics[width=0.425\textwidth]{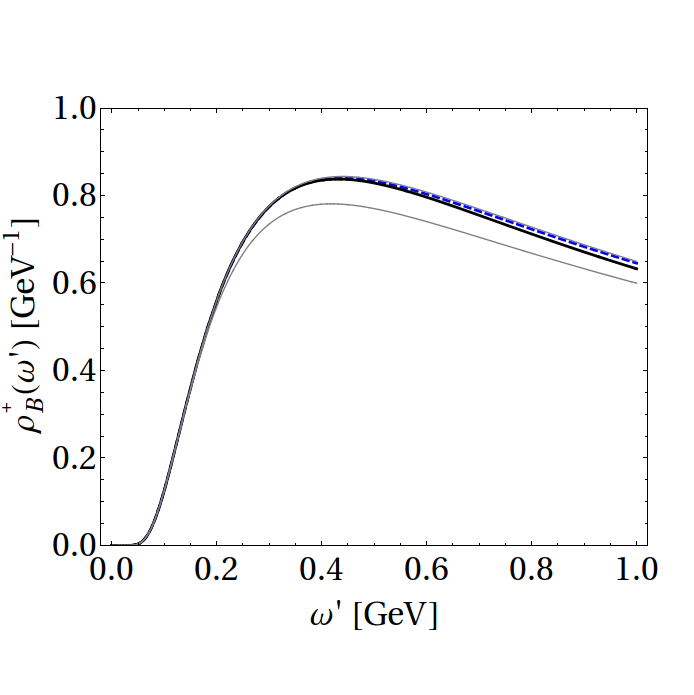} \quad 
 \includegraphics[width=0.425\textwidth]{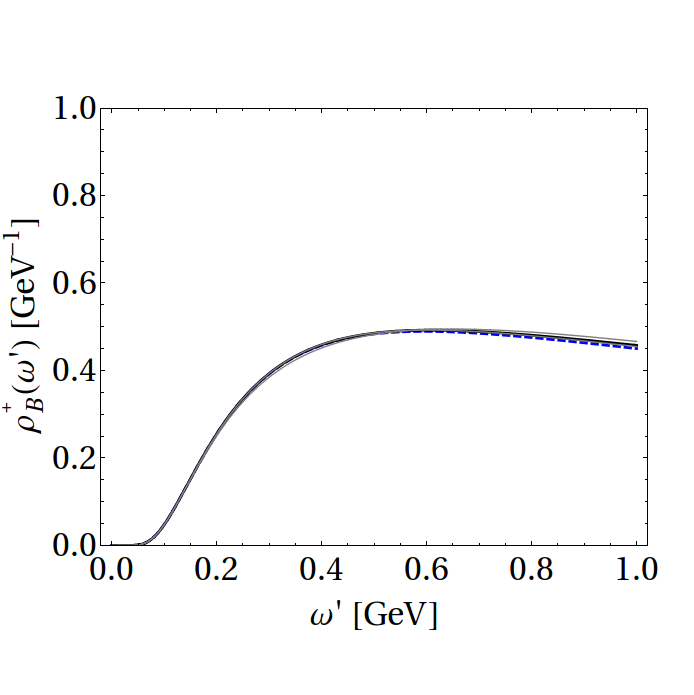} \\[-7mm]
 \includegraphics[width=0.425\textwidth]{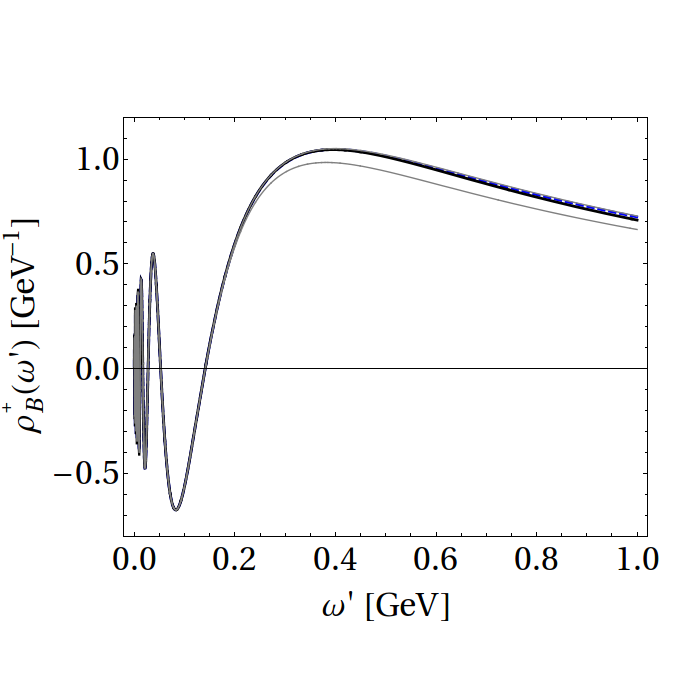} \quad 
 \includegraphics[width=0.425\textwidth]{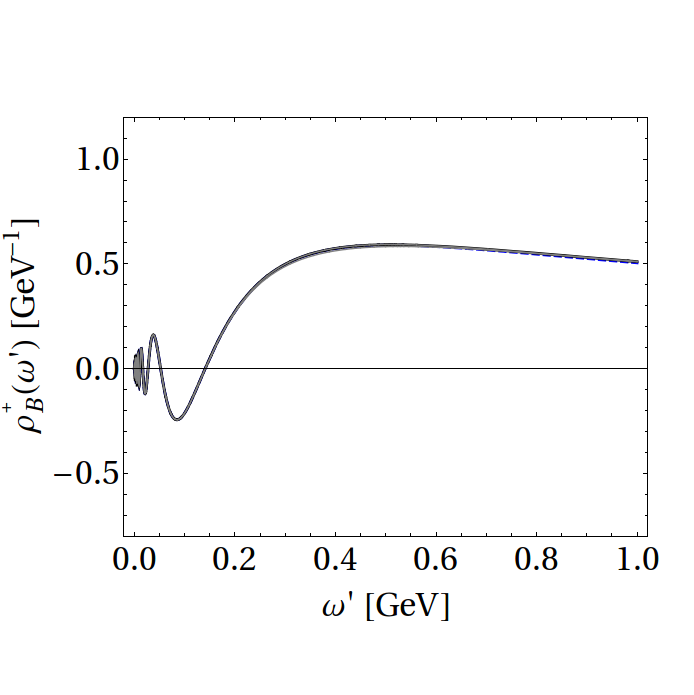} \\[-7mm]
 \includegraphics[width=0.425\textwidth]{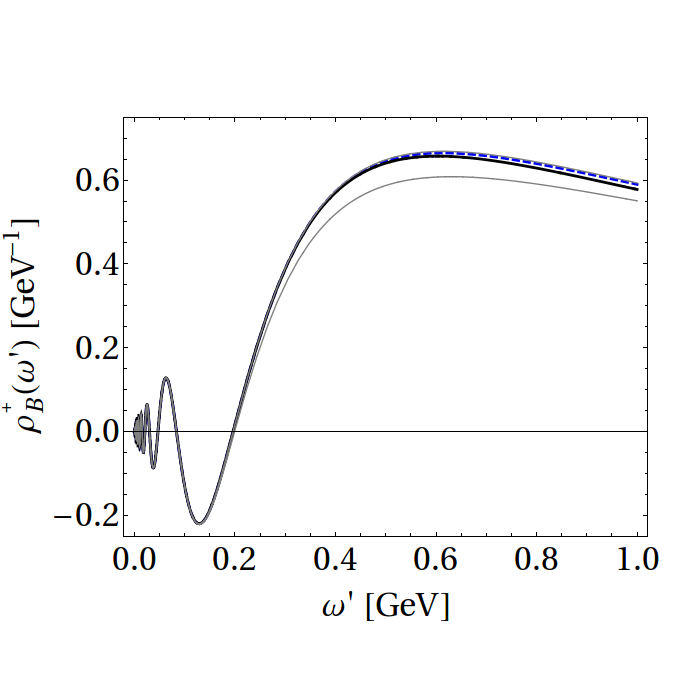} \quad 
 \includegraphics[width=0.425\textwidth]{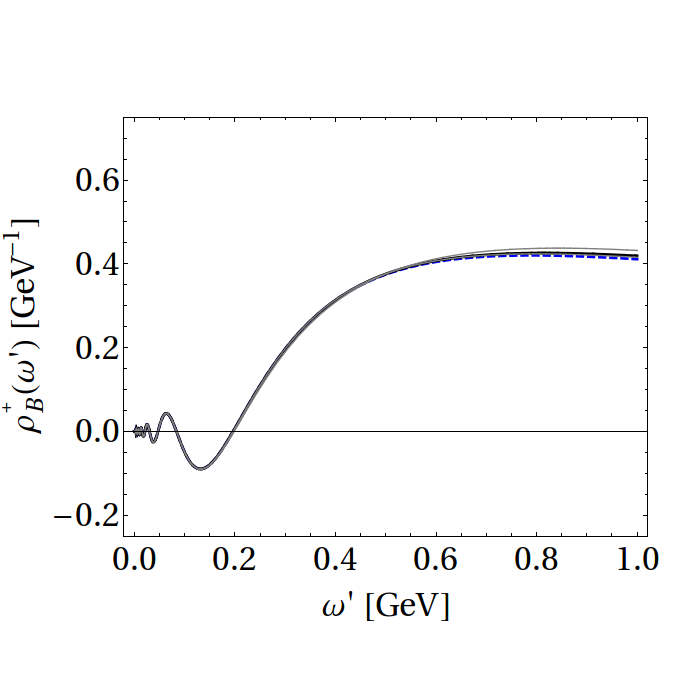}
 \caption{\label{fig:mod1_low} Behaviour of the dual function
   $\rho_B^+(\omega')$ at small values of $\omega'$. From top to
   bottom we show model-1, 2, 3. Left: For $\mu=\mu_0$. Right: For
   $\mu=5$~GeV. The solid line represents the result of the
   RG-improved parametrization (\ref{ansatz}). The grey
   lines indicate the variation of the parameter $\Omega$, appearing
   in the functions $p_n(\omega')$, within a factor of two. The thick
   (blue) dashed line refers to the original model functions $\rho^{\rm
     model-1,2,3}(\omega')$. (Notice that in the right plots, the model
   functions are evolved from $\mu_0 \to 5$~GeV.)}
\end{figure*}

\clearpage 

\appendix

\parindent0pt


\section{Some Properties of Special Functions}
We use the completeness relation for Bessel functions in the form
\begin{align}
& \int_0^\infty \frac{d\omega'}{\omega'} \, \frac{1}{\omega'} \,
J_{n}\left( 2 \, \sqrt{\frac{a}{\omega'} }\right)
   J_{n}\left( 2 \, \sqrt{\frac{b}{\omega'}}\right)
   = 
   \int_0^\infty d\omega \, J_{n}\left( 2 \, \sqrt{a \omega }\right)
   J_{n}\left( 2 \, \sqrt{ b  \omega}\right)
=\delta(a-b)\,.
   \label{complete}
\end{align}
Among others, this allows to revert the relation between
the regularized moments $M_{0,1}$ and the dual function
$\rho_B^+(\omega',\mu)$ 
for large values of $\hat\omega' \sim \mu$
 \begin{align}
 & \rho_B^+(\omega')  = \frac{1}{\omega'} 
  \int\limits_0^\infty \frac{d\Lambda_{\rm UV}}{\Lambda_{\rm UV}} \Bigg\{
   M_0 \Big|_{\bar\Lambda \to 0}
  \times  J_2\left(2 \, \sqrt{\frac{\Lambda_{\rm UV}}{\omega'}} \right) 
 +  \frac{\partial}{\partial \bar\Lambda} 
  \bigg( 2 M_0 - \frac{3 M_1}{\Lambda_{\rm UV}} 
  \bigg)_{\bar\Lambda \to 0} \times  \bar\Lambda \, 
  J_4\left(2 \, \sqrt{\frac{\Lambda_{\rm UV}}{\omega'}} \right)
+ \ldots \ 
   \Bigg\} \,.
 \end{align}

The Laguerre polynomials satisfy the orthogonality relation
\begin{align}
\int_0^\infty dz \, e^{-z} \, {\rm L}_n(z) \, {\rm L}_m(z) &= \delta_{nm} \,.
 \label{ortho_Laguerre}
\end{align}
The explicit expressions for $n=0,1,2$ read
\begin{align}
 L_0(z) &= 1 \,, \qquad L_1(z)=1-z \,, \qquad L_2(z) = \frac{1}{2} \left(z^2-4 z+2\right) \,.
\end{align}


\section{Explicit Form of RG Functions}

The expansions of the QCD $\beta$-function and the anomalous dimensions
in the LN kernel are defined as
\begin{align}
 \beta &= -8\pi \left( \beta_0 \left( \frac{\alpha_s}{4\pi} \right)^2 
  +\beta_1  \left( \frac{\alpha_s}{4\pi} \right)^3 
  +\beta_2  \left( \frac{\alpha_s}{4\pi} \right)^4 +\ldots \right)\,,
\end{align}
with
\begin{align}
 & \beta_0=\frac{11}{3} \, C_A - \frac{2}{3} \, n_f \,, 
 \quad \beta_1 = \frac{34}{3} \, C_A^2 - \frac{10}{3} \, C_A n_f - 2 \, C_F n_f \,, \cr 
 & \beta_2 = \frac{2857}{54}\,C_A^3
 + \left( C_F^2 - \frac{205}{18}\,C_F C_A
   - \frac{1415}{54}\,C_A^2 \right) n_f
 + \left( \frac{11}{9}\,C_F+\frac{79}{54}\,C_A\right)n_f^2
 \,.
\end{align}
With this we express the 3-loop running coupling constant as
\begin{align}
\alpha_s(\mu) &= \frac{2 \pi}{\beta _0
   L}  \left(1
   -\frac{\beta _1 \ln (2 L)}{2 \beta _0^2 L}+
   \frac{\beta_1^2 \left(\frac{\beta _0 \beta _2}{\beta
   _1^2}+\left(\ln (2 L)-\frac{1}{2}\right)^2-\frac{5}{4}\right)}{4 \beta
   _0^4 L^2}
   \right) 
   \,, \qquad 
   L=\ln \frac{\mu}{\Lambda_{\rm QCD}^{(n_f)}} \,.
   \label{as3}
\end{align}
For the considered range of RG scales, we take $n_f=4$ with $\Lambda_{\rm QCD}^{(4)}=299$~MeV.

The first coefficient of the anomalous dimension,
\begin{align}
 \gamma_+ = \frac{\alpha_s}{4\pi} \, \gamma_0 + \ldots \,,
\end{align}
reads $\gamma_0 = - 2\,C_F$ \cite{Lange:2003ff}.

The coefficients in the perturbative expansion of the 
cusp anomalous dimension,
\begin{align}
 \Gamma_{\rm cusp} = \frac{\alpha_s}{4\pi} \, \Gamma_0 +  \left( \frac{\alpha_s}{4\pi} \right)^2 \Gamma_1 +
\left( \frac{\alpha_s}{4\pi} \right)^3 \Gamma_2 + \ldots
  \,, 
\end{align}
read \cite{Korchemsky:1987wg,Korchemskaya:1992je,Moch:2004pa}
\begin{align}
 & \Gamma_0 = 4 \, C_F \,, \quad \Gamma_1 = C_F \left( 
 \left( \frac{268}{9} - \frac{4\pi^2}{3} \right) C_A - \frac{40}{9} \, n_f \right)
 \,, \qquad 
 \cr 
 & \Gamma_2 = 16 C_F \left(\left(-\frac{7 \zeta (3)}{3}-\frac{209}{108}+\frac{5 \pi
   ^2}{27}\right) C_A n_f+\left(\frac{11 \zeta
   (3)}{6}+\frac{245}{24}-\frac{67 \pi ^2}{54}+\frac{11 \pi ^4}{180}\right)
   C_A^2+\left(2 \zeta (3)-\frac{55}{24}\right) C_F
   n_f-\frac{n_f^2}{27}\right)
 \,.
 \cr &
\end{align}
The RG elements $U_{\omega'}(\mu,\mu_0)$ have been  defined in (\ref{BmesonRGE1}) as
\begin{align}
 U_{\omega'}(\mu,\mu_0) &= \exp\left[V(\mu,\mu_0) - g(\mu,\mu_0) \, \ln \frac{\mu_0}{\hat\omega'} \right] \,,
\end{align}
with the  RG functions $g$ and $V$ defined as (see e.g.\ \cite{Neubert:2004dd})
\begin{align}
g(\mu,\mu_0)= \int\limits_{\alpha_s(\mu_0)}^{\alpha_s(\mu)} 
 \frac{d\alpha}{\beta(\alpha)} \, \Gamma_{\rm cusp}(\alpha) 
\,, \qquad 
V(\mu,\mu_0) &= -\int\limits_{\alpha_s(\mu_0)}^{\alpha_s(\mu)} 
 \frac{d\alpha}{\beta(\alpha)} \left[ \gamma_+(\alpha) + \Gamma_{\rm cusp}(\alpha) 
 \int\limits_{\alpha_s(\mu_0)}^{\alpha} \frac{d\alpha'}{\beta(\alpha')} \right] \,.
\end{align}
In order to make the composition rule (\ref{comp}) manifest, it is convenient
to introduce a reference scale, $\mu_\ast$, which in the numerical analysis we
identify with the hadronic input scale $\mu_0$.
To this end, we rewrite the function $V$ as
\begin{align}
 V(\mu,\mu_0) &= -\int\limits_{\alpha_s(\mu_0)}^{\alpha_s(\mu)} 
 \frac{d\alpha}{\beta(\alpha)} \left[ \gamma_+(\alpha) + \Gamma_{\rm cusp}(\alpha) 
 \int\limits_{\alpha_s(\mu_\ast)}^{\alpha} \frac{d\alpha'}{\beta(\alpha')} \right]
 -\int\limits_{\alpha_s(\mu_0)}^{\alpha_s(\mu)} 
 \frac{d\alpha}{\beta(\alpha)} \, \Gamma_{\rm cusp}(\alpha) 
 \int\limits_{\alpha_s(\mu_0)}^{\alpha_s(\mu_\ast)} \frac{d\alpha'}{\beta(\alpha')} 
 \cr & \equiv V_\ast(\mu,\mu_0) - g(\mu,\mu_0) \, \ln \frac{\mu_\ast}{\mu_0} \,.
\end{align}
With this, the RG elements read
\begin{align}
 U_{\omega'}(\mu,\mu_0) &= \exp\left[V_\ast(\mu,\mu_0) - g(\mu,\mu_0) \, \ln \frac{\mu_\ast}{\hat\omega'} \right] \,,
\end{align}
and, by definition, 
\begin{align}
 V_\ast(\mu_2,\mu_1) + V_\ast(\mu_1,\mu_0) &= V_\ast(\mu_2,\mu_0) \,, \qquad 
 g(\mu_2,\mu_1) + g(\mu_1,\mu_0) = g(\mu_2,\mu_0) \,,
\end{align}
which implies the composition rule for $U_{\omega'}$.
Explicit expansions for the RG functions can now be obtained 
by inserting the perturbative expressions for the $\beta$-function
and anomalous dimension. Using the abbreviations 
\begin{align}
r_0  \equiv \frac{\alpha_s(\mu_0)}{\alpha_s(\mu_\ast)} \,, \qquad 
r_1 \equiv \frac{\alpha_s(\mu_1)}{\alpha_s(\mu_\ast)} \,,
\end{align}
we find $g(\mu_1,\mu_0)$ to 3-loop accuracy,
\begin{align}
 g(\mu_1,\mu_0) 
 &= \frac{\Gamma_0}{2\beta_0} \, \ln \frac{r_0}{r_1} 
  + \frac{\alpha_s(\mu_\ast)}{4\pi}  \, \frac{\beta_0 \Gamma_1 - \beta_1\Gamma_0}{2\beta_0^2}\, (r_0-r_1)
\cr & \qquad + 
\left(\frac{\alpha_s(\mu_\ast)}{4\pi} \right)^2 \frac{\beta _0^2 \Gamma _2-\beta _2 \beta _0 \Gamma _0-\beta _1 \beta _0
   \Gamma _1+\beta _1^2 \Gamma _0}{4 \beta _0^3} \, (r_0^2-r_1^2)
+\ldots 
\end{align}
Similarly, for the function $V_\ast(\mu_1,\mu_0)$ one gets
\begin{align}
 V_\ast(\mu_1,\mu_0)
 &= \frac{\pi}{\alpha_s(\mu_\ast)} \, \frac{\Gamma_0 }{\beta_0^2} 
 \left( \frac{1}{r_0} - \frac{1}{r_1} +\ln \frac{r_0}{r_1} \right)- \frac{\gamma_0}{2\beta_0} \, \ln \frac{r_0}{r_1} 
 \cr & \qquad 
  - \frac{\beta_0 \Gamma_1 - \beta_1\Gamma_0}{4\beta_0^3} \left( r_1-r_0 + \ln \frac{r_0}{r_1} \right)
  - \frac{\beta_1 \Gamma_0}{8\beta_0^3} \,(\ln^2 r_0 - \ln^2 r_1)
 + \ldots 
\end{align}
where we neglected terms of order $\alpha_s$, 
as the 2-loop result for the anomalous-dimension coefficient $\gamma_1$,
which will enter at that order,
is currently unknown. Notice that the so constructed expansions of $g$ and $V_\ast$,
and thus the expansion of $U_{\omega'}$, 
respect the composition rule (which would not have been the case if one
had expanded the function $V$ directly). \\[31mm]
\end{widetext}


\end{document}